\documentclass{article}
\usepackage{spconf,amsmath,graphicx,hyperref}
\usepackage{booktabs}
\usepackage{array}
\usepackage{pgfplots}
\pgfplotsset{compat=1.17}

\title{IS BROADER BETTER? A CONTROLLED STUDY OF MULTILINGUAL COVERAGE AND
PRETRAINING OBJECTIVE IN FROZEN SSL ENCODERS FOR SPEECH DEEPFAKE DETECTION}

\name{Benjamin Hurt$^{\star}$ \qquad Oscar O'Donnell$^{\dagger}$}
\address{$^{\star}$ University of California, Santa Barbara \\
         $^{\dagger}$ Columbia University}

\begin{document}
\ninept
\maketitle

\begin{abstract}
Frozen self-supervised (SSL) speech encoders are strong, low-cost front ends
for audio deepfake detection, and recent comparisons agree that large,
multilingual, discriminative encoders generalize best out of domain. These
comparisons fail to control for encoder capacity, pretraining objective, and multilingual
coverage together, identifying \emph{which} encoder wins without isolating
\emph{why}. We present a controlled decomposition with a fixed pipeline and trainable
capacity. We vary multilingual coverage on four wav2vec2-family
encoders, matched to ${\sim}315$M parameters. We isolate the pretraining objective on two encoders
matched on identical data. Coverage does not help monotonically, as out-of-domain
error drops sharply at the ${\sim}100$-language scale (XLS-R) but does not improve
further at the 1406-language extreme (MMS). We find that a mid-coverage encoder is strongest on farther out-of-domain sets, matching or surpassing a 577M-parameter model at 315M.
Its lead on these far sets, statistically significant under paired bootstrap, and on the official ASVspoof 5 cost metric holds under two backends.
Separately, masked-prediction pretraining generalizes better than contrastive on identical data (In-the-Wild EER 26.5\% vs. 46.8\%). Within this
fixed frozen-encoder recipe, we find that broader and larger models are not reliably better.
\end{abstract}


\section{Introduction}
\label{sec:intro}

Self-supervised learning (SSL) speech encoders are now the dominant front end
for audio deepfake detection~\cite{tak}, and recent work shows that even \emph{frozen}
encoders paired with a lightweight backend generalize well to out-of-domain
(OOD) attacks at a fraction of the training cost~\cite{tvot,spoofsuperb}. This makes the choice of frozen encoder a
fundamental design decision: the representation is fixed, so whatever it fails to
capture cannot be recovered downstream or in deployment. The property of an encoder's
pretraining that drives OOD generalization remains unresolved.

Recent comparisons report that large, multilingual, discriminative encoders
such as XLS-R generalize best \cite{spoofsuperb,phukan}, while concurrent studies question whether scale alone explains this~\cite{serrano,raptor}. But these comparisons vary
encoder capacity, pretraining objective, and multilingual coverage
\emph{simultaneously} (Spoof-SUPERB spans ${\sim}$4M--317M parameters~\cite{spoofsuperb}, and MMS is evaluated at 1B~\cite{phukan,serrano}), so they establish \emph{which}
encoder wins without isolating \emph{why}. The advantage could reflect scale,
objective, coverage, or their interaction. The confound is unresolved because prior comparisons rarely hold the other factors fixed.

To our knowledge, we provide the first controlled decomposition of coverage and objective at matched capacity. Fixing trainable-backend
capacity, pretraining objective, and pipeline, we sweep multilingual
coverage across four wav2vec2-family encoders matched to ${\sim}315$M
parameters (monolingual wav2vec2-LV60, 53-language XLSR-53, 128-language
XLS-R, and 1406-language MMS-300M),
with coverage, and its correlated data volume, as the varying factor. Our central finding is that
coverage does not help monotonically as
out-of-domain error drops sharply once coverage reaches the ${\sim}100$-language
scale (XLS-R), and does not improve at the 1406-language extreme (MMS). The 128-language encoder is the strongest choice on the farther out-of-domain
sets while both endpoints are worse. XLS-R is also best on the official ASVspoof~5 cost metric under two
independent backends (0.49--0.56, vs.\ 0.60--0.65 for MMS and $\geq$0.87 for
XEUS and the English-only encoders). At 315M parameters, it also matches or outperforms the 577M-parameter XEUS out of domain as one data point against parameter scale, rather than pretraining coverage, as the driver.
We further isolate the pretraining \emph{objective}, a factor prior comparisons did not isolate, by contrasting two encoders trained on
\emph{identical} data (wav2vec2 vs.\ HuBERT, both on Libri-Light-60k, both
${\sim}315$M): masked prediction generalizes markedly better than the
contrastive objective at fixed (monolingual) coverage.
Together, these controlled comparisons indicate that OOD generalization depends more on an encoder's pretraining --- its multilingual coverage and objective --- than on its parameter count, and that broader coverage is not always better.

Our contributions are: (i) to our knowledge, the first matched-capacity sweep of multilingual coverage across multiple coverage levels (1 to 1406 languages) for frozen SSL anti-spoofing, revealing a non-monotonic coverage effect that peaks near 100 languages; (ii) a matched-data isolation of pretraining objective, finding masked prediction generalizes better out of domain than contrastive learning; and (iii) multi-seed evaluation with paired-bootstrap significance testing across five benchmarks and two backends, including the official ASVspoof~5 cost metric, in contrast to prior single-run comparisons.

\section{Related Work}
\label{sec:related}

SSL front ends underpin most anti-spoofing systems. Early work established
their value with fine-tuned encoders \cite{wangyama,tak}, and
recent systems show frozen encoders with lightweight backends generalize well
at far lower cost \cite{tvot}. A parallel line compares encoders to identify
which generalize best. Spoof-SUPERB \cite{spoofsuperb} benchmarks 20 frozen SSL
models under a unified protocol and finds large multilingual discriminative
encoders strongest, attributing this to a bundle of scale, multilingual
pretraining, and speaker-aware objectives; concurrent preprints question whether scale alone explains robustness \cite{serrano,raptor}. Chetia Phukan et al.\ \cite{phukan} likewise
find multilingual encoders outperform monolingual ones, but compare 1B-parameter
multilingual models (XLS-R, MMS) against sub-350M monolingual ones, conflating
coverage with a ${\sim}3$--$10\times$ capacity gap, and treat multilinguality as
binary, leaving open whether \emph{more} coverage continues to help.
Notably, when trained on ASVspoof~2019 and tested on In-the-Wild, as in our
setting, MMS outperforms XLS-R in their cross-corpus evaluation, opposite to
our matched-capacity result; their use of 1B checkpoints, final-layer features,
and a different downstream pipeline may account for the discrepancy. These
comparisons, in general, vary encoder capacity, objective, and coverage
together (\cite{phukan} and \cite{serrano} evaluate MMS at 1B, while Spoof-SUPERB spans ${\sim}$4M--317M parameters), establishing ``winning'' encoders without isolating why.
RAPTOR \cite{raptor}, a controlled study at ${\sim}100$M parameters, similarly
attributes robustness to multilingual pretraining rather than scale, convergent with our findings at a different capacity band.
We provide the matched-capacity
and matched-data controls these comparisons lack. By sweeping coverage at fixed capacity, we directly test whether a broader model is better.

\section{Method}
\label{sec:method}

\subsection{Frozen-encoder pipeline}
\label{ssec:pipeline}
We have a fixed detection pipeline across all conditions, isolating the frozen encoder as the only variable. Each input waveform is passed through a frozen
SSL encoder.  We take the mean of the encoder's last four transformer layers as the
frame-level representation, following evidence that SSL layers encode
different information~\cite{pasad2021} and that spoofing cues
vary by layer~\cite{elkheir2025layer}.\footnote{For XEUS (19 blocks) we average blocks
16--19, matching the last-four convention.}
In parallel, a constant-Q cepstral
(CQCC) branch provides a spectral stream. Both streams are projected to $D{=}128$
and fused by the cross-attention recipe of \cite{tvot}, followed by a backend
(Section~\ref{ssec:cap}) and a linear bona-fide/spoof classifier. Only the
projections, fusion, backend, and classifier are trained. The encoder is frozen
throughout. Fusion uses 4-head cross-attention (dropout 0.1) while the MHFA backend
uses $K{=}64$ compression, $H{=}8$ heads, and 128-dim embeddings (dropout 0.2).
All systems are trained on ASVspoof~2019 LA with no augmentation, over three
seeds $\{42,1234,2025\}$, selecting the checkpoint by best LA19-dev EER. We use
AdamW with two learning rates (3\,$\times$\,10$^{-4}$ for the backend,
1\,$\times$\,10$^{-4}$ for the fusion front end), weight decay 10$^{-4}$, gradient
clipping 1.0, batch size 32, StepLR (halved every 10 epochs), and a
weighted-cross-entropy loss (bona-fide upweighted) with label smoothing 0.1. MHFA
cells train for 30 epochs; AASIST cells for 40, warm-started from the matched MHFA
head. The CQCC branch uses 60-dim features (20 coefficients with
$\Delta,\Delta\Delta$).

\subsection{Scale normalization as a control}
\label{ssec:norm}
Raw SSL features differ by roughly two orders of magnitude across
encoders (RMS ${\approx}770$ for XLSR-53 vs.\ ${\approx}5$ for MMS). Fed
directly to a projection with fixed initialization and learning rate, this
makes each encoder's first layer train at a ${\sim}100\times$ different
\emph{effective} learning rate, confounding encoder identity with input scale,
precisely the factor our study must control. We therefore apply a scale-matching
normalization to the frozen features before projection: a non-affine LayerNorm
followed by a fixed scalar $s$, identical for every encoder. We set $s{\approx}700$
to match the mean pre-normalization SSL magnitude, preserving the fusion's
original SSL-to-spectral balance. $s$ is a scale-matching constant, \emph{not} a
tuned hyperparameter. This removes cross-encoder scale confounding. As a validation that the normalization behaves as intended, it also resolves a bimodal in-domain instability we observe for XLSR-53 (seed std 16.68 $\rightarrow$ 1.63). Crucially, because the non-affine LayerNorm maps every encoder to unit variance regardless of
$s$, the \emph{cross-encoder} scale matching, our actual fairness mechanism, is
exact for any $s{>}0$. $s$ affects only the shared SSL-to-spectral branch balance,
which we fix to the anchor recipe's. The encoder comparison is therefore robust to
the choice of $s$ by construction, not by tuning.

\subsection{Two controlled axes}
\label{ssec:axes}
We isolate two pretraining factors by varying each while holding the other fixed.

\noindent\textbf{Coverage.} Four wav2vec2-family encoders (monolingual
wav2vec2-LV60 \cite{wav2vec2}, XLSR-53 \cite{xlsr53} (53 languages), XLS-R
\cite{xlsr} (128), and MMS-300M \cite{mms} (1406)) share the same contrastive
objective, architecture (24 layers, 1024-dim), and ${\sim}315$M parameters, and differ primarily in multilingual coverage. One caveat is irreducible, however, as coverage co-varies with pretraining data
\emph{volume} (${\approx}50$K to ${\approx}500$K hours) across public
checkpoints, which cannot be separated without controlled pretraining. We
therefore attribute effects to coverage and its correlated data scale, not to
language count alone.

\noindent\textbf{Objective.} wav2vec2-LV60 and HuBERT-LARGE \cite{hubert} are
both pretrained on \emph{identical} Libri-Light-60k data at ${\sim}315$M
parameters, differing in objective (contrastive vs.\ masked prediction). The
pretraining data is matched exactly, making this a clean contrast with respect to data.

\subsection{Capacity, reference, backends, metrics}
\label{ssec:cap}
All coverage and objective encoders are matched at 315.4M
parameters.\footnote{Exact counts: wav2vec2-LV60, XLSR-53, XLS-R, and MMS-300M
are 315.4M each (24 transformer layers, 1024-dim); HuBERT-LARGE 315.4M; WavLM-LARGE
315.5M. XEUS is 577M.} We include
the 577M massively-multilingual XEUS \cite{xeus} as a labeled
\emph{reference}, a larger, broader encoder outside the matched set, to ask
whether matched-capacity coverage can rival it. We evaluate two backends, MHFA
\cite{mhfa} and AASIST \cite{aasist}, on five benchmarks, each of increasing distance
from training: LA19 (in-domain) \cite{asvspoof2019}, LA21, DF21
\cite{asvspoof2021}, In-the-Wild (ITW) \cite{itw}, and ASVspoof~5 (ASV5)
\cite{asvspoof5}.
We report pooled EER as the mean over three seeds (seed std where relevant) and the
official ASVspoof~5 countermeasure min-DCF ($P_\text{spoof}{=}0.05$) on the
large-scale set. We build on the cross-attention SSL/spectral fusion recipe of
\cite{tvot}.

\section{Results}
\label{sec:results}

Table~\ref{tab:main} reports pooled EER and ASVspoof~5 min-DCF for both
controlled axes under the MHFA backend, with the 577M XEUS reference.

\begin{table}[t]
\centering
\caption{%
Pooled EER (\%, mean over 3 seeds) and official ASVspoof~5
min-DCF, MHFA backend. Coverage axis (matched ${\sim}315$M, wav2vec2 objective)
and objective axis (matched Libri-Light-60k data). XEUS is a 577M reference
outside the matched set. \textbf{Bold}: best per column among matched encoders.
$^{\ddagger}$WavLM adds a denoising objective \emph{and} more/broader data (94k h),
so it is not a clean same-data contrast.}
\label{tab:main}
\setlength{\tabcolsep}{3.2pt}
\footnotesize
\begin{tabular}{@{}l ccccc c@{}}
\toprule
\textbf{Encoder} & \textbf{LA19} & \textbf{LA21} & \textbf{DF21} & \textbf{ITW}
& \textbf{ASV5} & \textbf{mDCF}\\
\midrule
\multicolumn{7}{@{}l}{\emph{Coverage (matched ${\sim}315$M, contrastive)}}\\
wav2vec2-LV60 (1)      & 3.52 & 27.15 & 18.13 & 46.77 & 41.88 & 0.94\\
XLSR-53 (53)           & 3.67 & 30.32 & 22.90 & 38.97 & 25.40 & 0.65\\
XLS-R (128)            & 1.41 & \textbf{8.50} & \textbf{8.34}
                       & \textbf{18.30} & \textbf{21.18} & \textbf{0.56}\\
MMS-300M (1406)        & 1.43 & 11.19 & 8.48 & 22.71 & 24.36 & 0.65\\
\midrule
\multicolumn{7}{@{}l}{\emph{Objective (matched LL-60k data, ${\sim}315$M)}}\\
wav2vec2-LV60 (contr.) & 3.52 & 27.15 & 18.13 & 46.77 & 41.88 & 0.94\\
HuBERT (masked)        & 1.09 & 13.88 & 10.96 & 26.54 & 38.40 & 1.00\\
WavLM (masked+denoise)$^{\ddagger}$ & \textbf{0.92} & 11.75 & 12.69 & 46.03 & 43.96 & 1.00\\
\midrule
\multicolumn{7}{@{}l}{\emph{Reference (577M, outside matched set)}}\\
XEUS                   & 0.66 & 9.04 & 13.83 & 30.30 & 33.96 & 0.97\\
\bottomrule
\end{tabular}
\end{table}

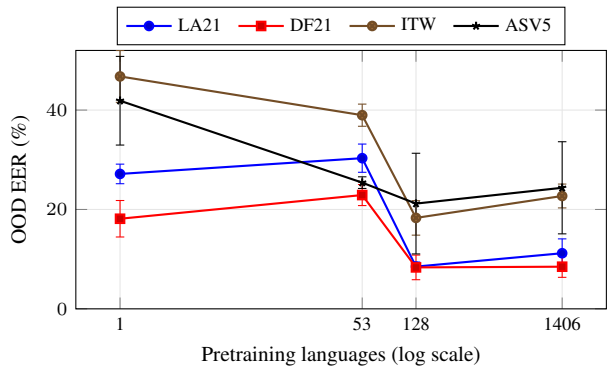
\begin{figure}[t]
\centering
\begin{tikzpicture}
\begin{axis}[
  width=\linewidth, height=5.0cm,
  xmode=log, log basis x=10,
  xtick={1,53,128,1406},
  xticklabels={1,53,128,1406},
  xlabel={Pretraining languages (log scale)},
  ylabel={OOD EER (\%)},
  ymin=0, ymax=52,
  xlabel style={font=\footnotesize}, ylabel style={font=\footnotesize},
  tick label style={font=\scriptsize},
  legend style={font=\scriptsize, at={(0.5,1.02)}, anchor=south,
                legend columns=4, /tikz/every even column/.append style={column sep=4pt}},
  grid=both, grid style={gray!20},
  mark size=1.6pt,
  error bars/y dir=both, error bars/y explicit,
]
\addplot+[thick] coordinates {(1,27.15)+-(0,1.97) (53,30.32)+-(0,2.84) (128,8.50)+-(0,0.92) (1406,11.19)+-(0,2.89)};
\addplot+[thick] coordinates {(1,18.13)+-(0,3.67) (53,22.90)+-(0,2.12) (128,8.34)+-(0,2.48) (1406,8.48)+-(0,2.14)};
\addplot+[thick] coordinates {(1,46.77)+-(0,5.21) (53,38.97)+-(0,2.23) (128,18.30)+-(0,3.47) (1406,22.71)+-(0,2.40)};
\addplot+[thick] coordinates {(1,41.88)+-(0,8.92) (53,25.40)+-(0,1.17) (128,21.18)+-(0,10.13) (1406,24.36)+-(0,9.27)};
\legend{LA21,DF21,ITW,ASV5}
\end{axis}
\end{tikzpicture}
\caption{Out-of-domain EER vs.\ multilingual coverage (MHFA, mean over 3 seeds,
error bars $\pm$1 std), for the four matched-capacity ($\sim$315M) wav2vec2-family
encoders. Every benchmark bottoms out at the 128-language XLS-R, then rises again
at MMS: coverage helps up to a point, not monotonically. Seed variance is large on
ASVspoof~5, where we rely on the more stable min-DCF (text).}
\label{fig:sweetspot}
\end{figure}

\subsection{Coverage helps up to a point}
\label{ssec:coverage}
With capacity and objective held fixed, OOD error is \emph{not} monotonic in multilingual coverage. The two higher-coverage encoders (128-language XLS-R and
1406-language MMS-300M) dramatically outperform both the monolingual endpoint
(wav2vec2-LV60) and the 53-language XLSR-53 on nearly every OOD benchmark by EER (the
cost-metric picture is qualified below). Among the two,
XLS-R has the lower EER on the two farther sets (ITW 18.30 vs.\ 22.71, ASV5 21.18
vs.\ 24.36) and on LA21 (8.50 vs.\ 11.19), and is level with MMS on DF21 (8.34
vs.\ 8.48). The trend is also non-monotonic at the low end: the 53-language XLSR-53 is \emph{worse} than the monolingual encoder on LA21 (30.32 vs.\ 27.15) and DF21 (22.90 vs.\ 18.13). OOD error therefore does not fall steadily with language count. It drops sharply once coverage reaches the $\sim$100-language scale (XLS-R) but shows no further gain at the 1406-language extreme, where it degrades by the cost metric below. Figure~\ref{fig:sweetspot} shows this shape directly, as every out-of-domain benchmark bottoms out at the 128-language XLS-R.

Because several encoders show high seed variance in EER on the large-scale
ASVspoof~5 set (XLS-R $21.18\pm10.13$, driven by one outlier seed of three), the
EER ranking there is seed-fragile, so we treat the more stable official ASVspoof~5
min-DCF as the primary metric on that set. On min-DCF, XLS-R attains the best cost
under \emph{both} backends (0.56 MHFA, 0.49 AASIST), ahead of MMS (0.65, 0.60) and
far ahead of the English-only encoders, whose min-DCF is near-degenerate on this
set (0.94--1.00). XLS-R's cost-metric lead under two architecturally
distinct backends, together with the consistent gap between high and low coverage
encoders, is the most stable form of the ranking. On ASV5
min-DCF, XLSR-53 and MMS are level (both 0.65), so the tier gap separates
XLS-R from the rest rather than splitting cleanly into two tiers.

The pattern also holds against a substantially larger model. Out of domain,
matched-capacity XLS-R (315M) matches or surpasses the 577M XEUS on the OOD
benchmarks (most sharply on ITW 18.30 vs.\ 30.30, ASVspoof~5 21.18 vs.\ 33.96,
and cost metric 0.56 vs.\ 0.97), despite roughly half the
encoder size. In domain
(LA19), XEUS is strongest (0.66 vs.\ 1.41), so the advantage is specifically one
of out-of-domain generalization rather than overall capability.

\noindent\textbf{Significance.} We test the key comparisons with a paired
bootstrap over evaluation utterances (class-stratified, 1000 replicates, MHFA;
the resampling unit is the utterance with the three trained models held fixed,
so intervals reflect evaluation-set sampling, not training-seed variance, which
we report separately). XLS-R's advantage over MMS is significant on LA21, ITW,
and ASV5 (95\% CIs excluding zero, all three seeds agreeing) but \emph{not} on
DF21 ($\Delta{=}{-}0.13$, CI $[-0.38,+0.13]$, MMS ahead on two of three seeds),
so the two higher-coverage encoders are statistically level on the near sets and
separated on the far ones (ITW $-4.42$ $[-4.81,-4.03]$; ASV5 $-3.19$
$[-3.30,-3.07]$). XLS-R's advantage over XEUS is significant on all four OOD sets
and largest on ITW ($-12.0$ $[-12.5,-11.6]$) and ASV5 ($-12.8$ $[-12.9,-12.7]$). On ASV5 the bootstrap confirms the
\emph{direction} of these gaps (all three seeds agree) but not their magnitude:
XLS-R's per-seed EER spans 15.0--32.9, so the utterance-level interval is much
narrower than the training-seed spread, and we rely on min-DCF for the ASV5
ranking accordingly.

\subsection{Pretraining objective matters independently}
\label{ssec:objective}
The objective contrast isolates a second factor. On \emph{identical}
Libri-Light-60k data at matched capacity, the masked-prediction encoder
(HuBERT) generalizes markedly better than the contrastive one (wav2vec2-LV60):
LA21 13.88 vs.\ 27.15, DF21 10.96 vs.\ 18.13, ITW 26.54 vs.\ 46.77, and
in-domain 1.09 vs.\ 3.52. Because pretraining data is matched exactly, this gap reflects the objective and its training recipe rather than data. Spoof-SUPERB shows the same ordering for this pair under an all-layer readout (ITW 21.0 vs.\ 40.5)~\cite{spoofsuperb}. Both English-only encoders remain weak on ASVspoof~5 (38.40 and 41.88 EER; near-degenerate min-DCF).
This finding is consistent with Section~\ref{ssec:coverage}: even on this English set, the multilingual encoders are far stronger, and monolingual pretraining transfers poorly regardless of objective.
Adding WavLM (masked prediction with a denoising objective, but trained on more
and broader data) does not cleanly extend the trend. It improves in-domain and
on LA21, while suffering on ITW and ASV5 (Table~\ref{tab:main}). Thus, a more
elaborate objective paired with broader data is not uniformly beneficial and,
unlike the wav2vec2/HuBERT pair, confounds objective with data.

\begin{table}[t]
\centering
\caption{AASIST backend: pooled EER (\%, mean over 3 seeds) and ASVspoof~5
min-DCF. XLS-R's cost-metric lead and the coverage-tier gap reproduce under a
second, architecturally distinct backend. XEUS shows a seed-dependent collapse
under AASIST (high variance on LA21/DF21). $^{\ddagger}$WavLM adds a denoising
objective and broader data, so it is not a clean same-data contrast.}
\label{tab:aasist}
\setlength{\tabcolsep}{3.2pt}
\footnotesize
\begin{tabular}{@{}l ccccc c@{}}
\toprule
\textbf{Encoder} & \textbf{LA19} & \textbf{LA21} & \textbf{DF21} & \textbf{ITW}
& \textbf{ASV5} & \textbf{mDCF}\\
\midrule
\multicolumn{7}{@{}l}{\emph{Coverage (matched ${\sim}315$M, contrastive)}}\\
wav2vec2-LV60 (1)   & 2.61 & 23.57 & 18.73 & 48.40 & 38.11 & 0.93\\
XLSR-53 (53)        & 6.00 & 29.27 & 19.83 & 39.61 & 33.58 & 0.93\\
XLS-R (128)         & 1.75 & 11.28 & 9.25 & \textbf{19.17}
                    & \textbf{17.08} & \textbf{0.49}\\
MMS-300M (1406)     & 2.30 & \textbf{9.32} & \textbf{7.76} & 23.75 & 20.97 & 0.60\\
\midrule
\multicolumn{7}{@{}l}{\emph{Objective (matched LL-60k data, ${\sim}315$M)}}\\
HuBERT (masked)     & 3.02 & 20.56 & 10.87 & 33.11 & 35.96 & 0.92\\
WavLM (masked+denoise)$^{\ddagger}$ & \textbf{1.51} & 21.82 & 17.95 & 52.30 & 40.94 & 0.89\\
\midrule
\multicolumn{7}{@{}l}{\emph{Reference (577M, outside matched set)}}\\
XEUS (577M ref)     & 2.95 & 21.87 & 24.11 & 40.72 & 35.05 & 0.87\\
\bottomrule
\end{tabular}
\end{table}

\subsection{What is and is not backend-robust}
\label{ssec:aasist}
Table~\ref{tab:aasist} repeats the study under AASIST, a graph-attention backend
architecturally unlike MHFA's attentive pooling, and we run the same paired
bootstrap on it. Several findings hold under both backends. The two higher-coverage encoders (XLS-R, MMS) far
outperform the monolingual and 53-language encoders on every OOD set. XLS-R
retains the best min-DCF (0.49 vs.\ MMS 0.60). And XLS-R's advantage over MMS on
the two \emph{far} sets is significant under both backends (ITW $-4.57$
$[-5.00,-4.12]$; ASV5 $-3.88$ $[-3.99,-3.79]$).

On the \emph{near} sets, however, the finer XLS-R-vs-MMS ordering does not merely
depend on the backend but significantly reverses. Under AASIST, MMS achieves lower
EER on both LA21 and DF21 (by $1.96$ and $1.49$ points, CIs excluding zero),
whereas MHFA favored XLS-R on LA21 and tied on DF21. Thus, the backend determines the winning high-coverage encoder on the near sets, and XLS-R's lead is robust
only on the far OOD sets and the cost metric. Against XEUS, XLS-R's advantage is significant on all four OOD sets under both backends (more markedly AASIST). AASIST severely destabilizes XEUS as one of three seeds collapses on every benchmark (e.g.\ LA21 40.6\%). Excluding the destabilized seed, the two encoders are closely matched on LA21, so the pooled gap is driven almost entirely by the collapse. We therefore treat MHFA as the primary basis for the XEUS comparison.

\subsection{Summary}
\label{ssec:summary}
The two controlled contrasts separate factors that prior comparisons entangled.
Out-of-domain generalization improves sharply with coverage up to the
${\sim}100$-language scale and then plateaus (pushing to 1406 languages, MMS,
does not consistently improve EER and worsens the cost metric), showing more coverage
is not reliably better. Performance improves under a masked-prediction rather than a
contrastive objective on matched data. Neither effect reduces to parameter count, as a 315M encoder matches or surpasses the 577M reference out of domain.

\section{Discussion}
\label{sec:discussion}

\noindent\textbf{Why does coverage peak?} Our design does not test the mechanism, but
the pattern is consistent with a trade-off between breadth and per-language
representation quality. XLS-R's 128-language pretraining plausibly supplies
enough phonetic and acoustic diversity to transfer to unseen attacks, while MMS
spreads comparable capacity across 1406 mostly low-resource languages, most
irrelevant to the English evaluation sets, diluting the
representations.
Consistent with this, MMS adds to XLS-R's pretraining corpora ${\sim}55$K hours
of largely single-speaker scripture readings, and its authors report slight
degradation relative to XLS-R on high-resource languages, including English,
in fine-tuned ASR~\cite{mms}. Isolating this mechanism would require
controlled pretraining, which public checkpoints do not permit.

\noindent\textbf{Limitations.} Three bound our claims. First, across public
checkpoints multilingual coverage co-varies with pretraining data \emph{volume}
(${\approx}50$K to ${\approx}500$K hours). We attribute effects to coverage and
its correlated data scale, not language count in isolation. Second, all systems
are trained on a single corpus (ASVspoof~2019 LA), so our conclusions are scoped to
this training regime and the fixed detection recipe. Third, XEUS is a single
larger reference on a different encoder lineage, not a matched cell. The
observation that a 315M encoder surpasses XEUS out of domain is one
data point against scale-as-driver, not a general scaling law.

\noindent\textbf{Pretraining/evaluation provenance.} Several evaluation
benchmarks share provenance with our encoders' pretraining corpora:
LA19/LA21/DF21 derive substantially from VCTK, which XEUS's pretraining
includes, and ASVspoof~5 is sourced from the English portion of Multilingual
LibriSpeech (MLS), a component of XLS-R's, XLSR-53's, and MMS's pretraining
mixtures. We cannot fully separate coverage-driven generalization from this
exposure on these benchmarks. Two facts bound, but do not fully resolve, the concern.
First, XLSR-53 is comparably MLS-exposed yet sits in the low-performing tier of our
coverage sweep, so MLS exposure alone does not predict the ranking.
Second, MMS-300M's pretraining includes the same corpora as XLS-R, MLS among
them \cite{mms}, yet XLS-R still leads it on ASVspoof~5, so shared MLS exposure
does not account for that margin. However, we cannot rule out that MLS exposure
inflates absolute ASVspoof~5 performance for the exposed encoders.

\section{Conclusion}
\label{sec:conclusion}

We presented a controlled decomposition of what drives out-of-domain generalization in frozen SSL encoders for audio deepfake detection. By isolating multilingual coverage at matched capacity and pretraining objective at matched data, we separated two factors that prior comparisons varied together. Within this fixed recipe, coverage helps only up to the ${\sim}100$-language scale rather than
monotonically: a 315M mid-coverage encoder (XLS-R) is the strongest out-of-domain
choice on the farther sets, while pushing coverage to 1406 languages does not
consistently help. Separately,
a masked-prediction objective generalizes better than a contrastive one on
identical data. For frozen anti-spoofing, encoder choice should target sufficient
but not maximal multilingual coverage and a masked-prediction objective rather
than maximal scale. XLS-R emerges as a strong matched-capacity default. Whether
these effects generalize across training corpora and detection recipes, and what
mechanism produces the coverage peak, are open questions for future work.

\section*{Acknowledgments}
This work was supported in part by computational resources provided by the NSF ACCESS program under allocation CIS261055. We thank James O'Brien at Sound Ethics for his mentorship. The authors used Claude (Anthropic) for feedback on drafts and suggested revisions to wording in the Abstract, Introduction, Related Work, and Discussion for clarity. All AI-assisted content was reviewed and verified by the authors, who take full responsibility for this manuscript.

\vfill\pagebreak

\section{Compliance with Ethical Standards}
This work studies detection of, not generation of, synthetic speech, using only
publicly available benchmark datasets (ASVspoof~2019/21/5, In-the-Wild) under
their respective licenses. No new speech was collected and no human subjects
were involved. We release no attack systems. All encoders are publicly available checkpoints
and all backends and training configurations are specified in
Section~\ref{sec:method}.

\bibliographystyle{IEEEbib}
\bibliography{refs}

\begin{thebibliography}{10}

\bibitem{tak}
Hemlata Tak, Massimiliano Todisco, Xin Wang, Jee weon Jung, Junichi Yamagishi, and Nicholas Evans,
\newblock ``Automatic speaker verification spoofing and deepfake detection using wav2vec 2.0 and data augmentation,''
\newblock in {\em Proc. Odyssey}, 2022.

\bibitem{tvot}
Yassine~El Kheir, Arnab Das, Enes~Erdem Erdogan, Fabian Ritter-Guttierez, Tim Polzehl, and Sebastian M\"oller,
\newblock ``Two views, one truth: Spectral and self-supervised features fusion for robust speech deepfake detection,''
\newblock in {\em Proc. IEEE Workshop Appl. Signal Process. Audio Acoust. (WASPAA)}, 2025.

\bibitem{spoofsuperb}
Hashim Ali, Nithin~Sai Adupa, Surya Subramani, and Hafiz Malik,
\newblock ``A {SUPERB}-style benchmark of self-supervised speech models for audio deepfake detection,''
\newblock in {\em Proc. IEEE ICASSP}, 2026.

\bibitem{phukan}
Orchid~Chetia Phukan, Gautam~Siddharth Kashyap, Arun~Balaji Buduru, and Rajesh Sharma,
\newblock ``Heterogeneity over homogeneity: Investigating multilingual speech pre-trained models for detecting audio deepfake,''
\newblock in {\em Findings of the Association for Computational Linguistics: NAACL}, 2024, pp. 2496--2506.

\bibitem{serrano}
Pierre Serrano, Rapha\"el Duroselle, Florian Angulo, Jean-Fran\c{c}ois Bonastre, and Olivier Boeffard,
\newblock ``Improving out-of-domain audio deepfake detection via layer selection and fusion of {SSL}-based countermeasures,''
\newblock {\em arXiv preprint arXiv:2509.12003}, 2025.

\bibitem{raptor}
Ajinkya Kulkarni, Sandipana Dowerah, Atharva Kulkarni, Tanel Alum\"ae, and Mathew Magimai-Doss,
\newblock ``Do compact {SSL} backbones matter for audio deepfake detection? a controlled study with {RAPTOR},''
\newblock {\em arXiv preprint arXiv:2603.06164}, 2026.

\bibitem{wangyama}
Xin Wang and Junichi Yamagishi,
\newblock ``Investigating self-supervised front ends for speech spoofing countermeasures,''
\newblock in {\em Proc. Odyssey}, 2022.

\bibitem{pasad2021}
Ankita Pasad, Ju-Chieh Chou, and Karen Livescu,
\newblock ``Layer-wise analysis of a self-supervised speech representation model,''
\newblock in {\em Proc. IEEE Automatic Speech Recognition and Understanding Workshop (ASRU)}, 2021, pp. 914--921.

\bibitem{elkheir2025layer}
Yassine~El Kheir, Younes Samih, Suraj Maharjan, Tim Polzehl, and Sebastian M\"oller,
\newblock ``Comprehensive layer-wise analysis of {SSL} models for audio deepfake detection,''
\newblock in {\em Findings of the Association for Computational Linguistics: NAACL}, 2025, pp. 4070--4082.

\bibitem{wav2vec2}
Alexei Baevski, Henry Zhou, Abdelrahman Mohamed, and Michael Auli,
\newblock ``wav2vec 2.0: A framework for self-supervised learning of speech representations,''
\newblock in {\em Adv. Neural Inf. Process. Syst. (NeurIPS)}, 2020.

\bibitem{xlsr53}
Alexis Conneau, Alexei Baevski, Ronan Collobert, Abdelrahman Mohamed, and Michael Auli,
\newblock ``Unsupervised cross-lingual representation learning for speech recognition,''
\newblock in {\em Proc. Interspeech}, 2021.

\bibitem{xlsr}
Arun Babu, Changhan Wang, Andros Tjandra, Kushal Lakhotia, Qiantong Xu, Naman Goyal, Kritika Singh, Patrick von Platen, Yatharth Saraf, Juan Pino, Alexei Baevski, Alexis Conneau, and Michael Auli,
\newblock ``{XLS-R}: Self-supervised cross-lingual speech representation learning at scale,''
\newblock in {\em Proc. Interspeech}, 2022, pp. 2278--2282.

\bibitem{mms}
Vineel Pratap, Andros Tjandra, Bowen Shi, Paden Tomasello, Arun Babu, Sayani Kundu, Ali Elkahky, Zhaoheng Ni, Apoorv Vyas, Maryam Fazel-Zarandi, Alexei Baevski, Yossi Adi, Xiaohui Zhang, Wei-Ning Hsu, Alexis Conneau, and Michael Auli,
\newblock ``Scaling speech technology to 1{,}000+ languages,''
\newblock {\em J. Mach. Learn. Res.}, vol. 25, no. 97, pp. 1--52, 2024.

\bibitem{hubert}
Wei-Ning Hsu, Benjamin Bolte, Yao-Hung~Hubert Tsai, Kushal Lakhotia, Ruslan Salakhutdinov, and Abdelrahman Mohamed,
\newblock ``{HuBERT}: Self-supervised speech representation learning by masked prediction of hidden units,''
\newblock {\em IEEE/ACM Trans. Audio, Speech, Language Process.}, vol. 29, pp. 3451--3460, 2021.

\bibitem{xeus}
William Chen, Wangyou Zhang, Yifan Peng, Xinjian Li, Jinchuan Tian, Jiatong Shi, Xuankai Chang, Soumi Maiti, Karen Livescu, and Shinji Watanabe,
\newblock ``Towards robust speech representation learning for thousands of languages,''
\newblock in {\em Proc. EMNLP}, 2024.

\bibitem{mhfa}
Junyi Peng, Old\v{r}ich Plchot, Themos Stafylakis, Ladislav Mo\v{s}ner, Luk\'a\v{s} Burget, and Jan \v{C}ernock\'y,
\newblock ``An attention-based backend allowing efficient fine-tuning of transformer models for speaker verification,''
\newblock in {\em Proc. IEEE Spoken Language Technology Workshop (SLT)}, 2022.

\bibitem{aasist}
Jee weon Jung, Hee-Soo Heo, Hemlata Tak, Hye jin Shim, Joon~Son Chung, Bong-Jin Lee, Ha-Jin Yu, and Nicholas Evans,
\newblock ``{AASIST}: Audio anti-spoofing using integrated spectro-temporal graph attention networks,''
\newblock in {\em Proc. IEEE ICASSP}, 2022.

\bibitem{asvspoof2019}
Xin Wang, Junichi Yamagishi, Massimiliano Todisco, et~al.,
\newblock ``{ASVspoof} 2019: A large-scale public database of synthesized, converted and replayed speech,''
\newblock {\em Computer Speech \& Language}, vol. 64, pp. 101114, 2020.

\bibitem{asvspoof2021}
Xuechen Liu, Xin Wang, Md~Sahidullah, Jose Patino, H\'ector Delgado, Tomi Kinnunen, Massimiliano Todisco, Junichi Yamagishi, Nicholas Evans, Andreas Nautsch, and Kong~Aik Lee,
\newblock ``{ASVspoof} 2021: Towards spoofed and deepfake speech detection in the wild,''
\newblock {\em IEEE/ACM Trans. Audio, Speech, Language Process.}, vol. 31, pp. 2507--2522, 2023.

\bibitem{itw}
Nicolas M\"uller, Pavel Czempin, Franziska Dieckmann, Adam Froghyar, and Konstantin B\"ottinger,
\newblock ``Does audio deepfake detection generalize?,''
\newblock in {\em Proc. Interspeech}, 2022.

\bibitem{asvspoof5}
Xin Wang, H\'ector Delgado, Hemlata Tak, Jee weon Jung, Hye jin Shim, Massimiliano Todisco, Ivan Kukanov, Xuechen Liu, Md~Sahidullah, Tomi Kinnunen, Nicholas Evans, Kong~Aik Lee, and Junichi Yamagishi,
\newblock ``{ASVspoof} 5: Crowdsourced speech data, deepfakes, and adversarial attacks at scale,''
\newblock in {\em Proc. ASVspoof Workshop}, 2024,
\newblock arXiv:2408.08739.

\end{thebibliography}

\end{document}